\documentclass[letterpaper, 10 pt, conference, onecolumn, draft]{ieeeconf}  

\IEEEoverridecommandlockouts                              
\usepackage{xcolor}
\usepackage{amsmath,amssymb,amsfonts}
\usepackage{wrapfig}
\usepackage{graphicx}
\usepackage{algorithm}
\usepackage{algcompatible}
\algnewcommand\algorithmicinput{\textbf{Input:}}
\algnewcommand\INPUT{\item[\algorithmicinput]}
\algnewcommand\algorithmoutput{\textbf{Output:}}
\algnewcommand\OUTPUT{\item[\algorithmoutput]}
\usepackage{hyperref}
\hypersetup{hidelinks}
\usepackage{textcomp}

\usepackage{balance}
\usepackage{subcaption}
\usepackage{placeins}
\usepackage[binary-units]{siunitx}

\definecolor{cblue}{HTML}{045275}
\definecolor{cred}{HTML}{F0746E}
\definecolor{cgreen}{HTML}{7CCBA2}
\definecolor{legreddraw}{HTML}{E41A1C}
\definecolor{legbluedraw}{HTML}{377EB8}
\definecolor{leggreendraw}{HTML}{4DAF4A}
\definecolor{legvioletdraw}{HTML}{984EA3}
\definecolor{legredfill}{HTML}{B61516}
\definecolor{legbluefill}{HTML}{2C6593}
\definecolor{leggreenfill}{HTML}{3E8C3B}
\definecolor{legvioletfill}{HTML}{7A3E82}

\usepackage{amsmath, amssymb, amsfonts, mathtools, bm}
\usepackage{stmaryrd} 
\usepackage{mleftright}
\usepackage{mymath} 
\usepackage{arydshln} 

\usepackage{parskip} 
\usepackage{tabularx,ragged2e,booktabs,caption,multirow} 
\usepackage{array}
\usepackage{makecell} 

\usepackage{extarrows}
\RequirePackage{rotating} 
\RequirePackage{fix-cm}
\usepackage{stackengine}
\usepackage{multicol}
\usepackage{rotating}
\usepackage{booktabs}

\usepackage{pgfplots}
\pgfplotsset{compat=1.16}
\usepgfplotslibrary{patchplots}
\usepgfplotslibrary{colormaps}
\pgfplotsset{compat=newest}
\pgfplotsset{
  layers/axis lines on top/.define layer set={
    axis background,
    axis grid,
    axis ticks,
    axis tick labels,
    pre main,
    main,
    axis lines,
    axis descriptions,
    axis foreground,
  }{/pgfplots/layers/standard},
}
\usepgfplotslibrary{fillbetween}
\usepgfplotslibrary{colorbrewer}

\usepackage{tikz}
\usetikzlibrary{intersections}
\usetikzlibrary{shapes, arrows, shapes.misc, arrows.meta, positioning, matrix, calc, fit, fadings, patterns}
\usetikzlibrary{calc,patterns,decorations.pathmorphing,decorations.markings,arrows, arrows.meta,pgfplots.colorbrewer}
\usetikzlibrary{shapes, arrows, arrows.meta, positioning, calc,pgfplots.colorbrewer}
\newlength{\Llong}
\newlength{\Lmid}
\newlength{\Lshort}
\usepackage{csvsimple}
\pgfdeclarelayer{bg}    
\pgfsetlayers{bg,main}  
\usetikzlibrary{plotmarks}
\newenvironment{customlegend}[1][]{%
    \begingroup
    \csname pgfplots@init@cleared@structures\endcsname
    \pgfplotsset{#1}%
}{%
    \csname pgfplots@createlegend\endcsname
    \endgroup
}%
\def\addlegendimage{\csname pgfplots@addlegendimage\endcsname}

\usepackage[normalem]{ulem}

\pgfplotsset{colormap={CMsurf}{color=(white) color=(legbluedraw) color=(legvioletdraw) color=(legreddraw) color=(yellow)}}

\pgfplotsset{colormap={CMsurfrev}{color=(yellow) color=(legreddraw) color=(legvioletdraw) color=(legbluedraw) color=(white)}}

\pgfplotsset{colormap={CMsurfr}{color=(yellow) color=(legreddraw) color=(legvioletdraw) color=(legbluedraw) color=(white)}}
\pgfplotsset{
    colormap/viridis,
}
\tikzset{
    png export/.style={
        external/system call/.add={}{; convert -density 150 "\image.pdf" -quality 90 "\image.png"},
        /pgf/images/external info,
        /pgf/images/include external/.code={
            \includegraphics[width=\pgfexternalwidth,height=\pgfexternalheight]{##1.png}
        },
    }
}

\title{\LARGE \bf
Closed-Loop Sensitivity Identification for Cross-Directional Systems
}

\author{Callum Umana Stuart${^\dagger}$ and Idris Kempf${^*}$
\thanks{Both authors are with the Department of Engineering Science, University of Oxford, Oxford, UK.}\\
\thanks{${^\dagger}$C. U. S. was supported by the Engineering Undergraduate Research Opportunities Programme (EUROP) scheme of the Department of Engineering Science, University of Oxford, Oxford, UK.}\\
\thanks{$^*$Corresponding author: {\tt\footnotesize{idris.kempf@eng.ox.ac.uk}}}}

\begin{document}

\maketitle
\thispagestyle{empty}
\pagestyle{empty}

\begin{abstract}
At Diamond Light Source, the UK's national synchrotron facility, electron beam disturbances are attenuated by the fast orbit feedback (FOFB), which controls a cross-directional (CD) system with hundreds of inputs and outputs. Due to the inability to measure the disturbances in real-time, the closed-loop sensitivity of the FOFB can only be evaluated indirectly, making it difficult to compare FOFB algorithms and detect faults. Existing methods rely on comparing open-loop with closed-loop measurements, but they are prone to instabilities and actuator saturation because of the system's strong directionality. Here, we introduce a reference signal to estimate the complementary sensitivity in closed loop. By decoupling the system into sets of single-input, single-output (SISO) systems, the reference signal is designed mode-by-mode, accommodating the system's strong directionality. Additionally, a lower bound on the reference amplitude is derived to limit the estimation error in the presence of disturbances and measurement noise. This method enables the use of SISO system identification techniques, making it suitable for large-scale systems. It not only facilitates performance estimation of ill-conditioned CD systems in closed-loop but also provides a signal for fault detection. The potential applications of this approach extend to other CD systems, such as papermaking, steel rolling, or battery manufacturing processes.
\end{abstract}

\section{INTRODUCTION}

Diamond Light Source (Diamond) is the UK's national synchrotron facility that produces synchrotron radiation for research. It is emitted by an electron beam circulating at relativistic speeds around the \emph{storage ring}. The synchrotron radiation spans the electromagnetic spectrum from infrared to X-rays and is used for various scientific techniques, such as microscopy, scattering, diffraction, and spectroscopy~\cite{WIEDEMANN}.

A critical factor in synchrotron performance is the brightness of the synchrotron radiation, which can be significantly impacted by disturbances of the electron beam. These disturbances are caused by electromagnetic radiation, girder and machine component vibrations, or by machine operations~\cite{IANDIAMONDIORBIT}. To attenuate these disturbances and minimise the beam trajectory error, the \emph{fast orbit feedback} (FOFB) system is employed that uses hundreds of \emph{corrector magnets} (inputs) and \emph{beam position monitors} (outputs) at a rate of $\SI{10}{\kHz}$. The dynamics of the electron beam are modelled by a cross-directional (CD) system~\cite{SANDIRAWINDUP}:
\begin{align}\label{eq:CDsystem}
y(s)=P(s)u(s)+d(s),
\end{align}
where $s\in\C$ is the Laplace variable, $\inR{R}{n_y}{n_u}$ the \emph{response matrix}, $P(s)\eqdef Rg(s)$ the plant, $g:\C\mapsto\C$ the scalar actuator dynamics, $u:\C\mapsto\C^{n_u}$ are the inputs, $y:\C\mapsto\C^{n_y}$ the outputs and $d:\C\mapsto\C^{n_y}$ the disturbances. The separation of the plant into a matrix of constant values and a scalar dynamic term allows~\eqref{eq:CDsystem} to be diagonalised using the singular value decomposition (SVD) $R=U\Sigma\trans{V}$, which is referred as the \emph{modal transformation}~\cite{HEATH}. 

At Diamond, the modal transformation is used with the \emph{internal model control} (IMC) structure from Fig.~\ref{fig:diagram}, where $\hat{P}(s)$ refers to the plant model. The IMC filter $Q:\C^{n_y}\mapsto\C^{n_u}$ is based on a standard approach~\cite{morari1989robust} and combines a pseudo-inverse $\pinv{R}$ with a scalar transfer function $q:\C\mapsto\C$ that (partially) inverts the actuator dynamics. For synchrotrons, the response matrix is ill-conditioned with condition numbers $\kappa(R)\eqdef\twonorm{R}/\twonorm{\pinv{R}}$ ranging from $10^3$ to $10^4$~\cite{SANDIRAWINDUP}, making~\eqref{eq:CDsystem} prone to actuator saturation and sensitive to modelling errors~\cite{ILLCONDPLANTS}. This is accounted for using the static pre-compensator $\inR{\Gamma}{n_y}{n_y}$. Other examples of large-scale CD systems can be found in process engineering, paper making, web processes, and metal rolling~\cite{CROSSDIR}.

For synchrotron operation, it is crucial that the FOFB meets the theoretical performance specifications, i.e.\ that the \emph{sensitivity} $S:\C^{n_y}\mapsto\C^{n_y}$ in $y(s)=S(s)d(s)$ has the expected gains. However, $d(s)$ cannot be measured when the FOFB is operational, prohibiting $S(s)$ to be estimated in closed-loop. To identify the estimate $\hat{S}(s)$, existing methods rely on comparing open-loop with closed-loop measurements. One approach is to compute $\twonorm{y^{\mathrm{cl}}(\jw)}/\twonorm{y^{\mathrm{ol}}(\jw)}$, where~\cite{SANDIRACONTROLDESIGN}
\begin{align*}
\frac{y^{\mathrm{cl}}(s)}{y^{\mathrm{ol}}(s)}\eqdef\frac{S(s)d(s) -T(s)n(s)}{d(s)+n(s)},
\end{align*}
and $n:\C\mapsto\C^{n_y}$ is the measurement noise and $T(s)\eqdef I-S(s)$ the \emph{complementary sensitivity}. However, the disturbance can be time-varying and the 2-norm inappropriate for systems with large $\kappa(R)$. 

Another approach is to add an input signal $r_u:\C\mapsto\C^{n_u}$ and measure the output in both open and closed loop~\cite{Brones:2023lqd}, so that 
\begin{align*}
\frac{y^{\mathrm{cl}}(s)}{y^{\mathrm{ol}}(s)}\eqdef\frac{S(s)P(s)r_u(s)+S(s)d(s)}{P(s)r_u(s)+d(s)}.
\end{align*}
Suppose that $r_u(s)=e_i \rho_i(s)$ with $e_i$ being the $i$th standard basis vector and $\rho_i :\C\mapsto\C$ a scalar function, then the ratio $y^{\mathrm{cl}}_i(s)/y^{\mathrm{ol}}_i(s)$ for output $i$ becomes
\begin{align*}
\frac{y^{\mathrm{cl}}_i(s)}{y^{\mathrm{ol}}_i(s)}=\frac{\sum_j S_{i,j}(s)\left( P_{j,i}(s)\rho_i(s)+d_j(s)\right)}{P_{i,i}(s)\rho_i(s)+d_i(s)}.
\end{align*}
For $\abs{P_{j,i}(\jw)\rho_i(\jw)}\gg\abs{d_j(\jw)}\,\,\forall j$, it holds that
\begin{align*}
\frac{\abs{y^{\mathrm{cl}}_i(\jw)}}{\abs{y^{\mathrm{ol}}_i(\jw)}}\approx \frac{\sum_j\abs{S_{i,j}(\jw)}\abs{P_{j,i}(\jw)}}{\abs{P_{i,i}(\jw)}},
\end{align*}
from which $S_{i,i}$ can be estimated if the system is diagonally dominant~\cite{FGOLDVARGA}, i.e. if $\abs{P_{i,i}(\jw)}\gg\abs{P_{j,i}(\jw)}\,\forall j\neq i$. However, the requirement $\abs{P_{j,i}(\jw)\rho_i(\jw)}\gg\abs{d_j(\jw)}$, i.e.\ a large $r_u(s)$, will produce inadmissibly large beam trajectory error in open-loop, in particular in direction of higher-order modes associated with small singular values of $R$. To reliably estimate the sensitivity, system~\eqref{eq:CDsystem} must therefore be operated in closed loop.

In this paper, we propose introducing an output reference signal $r:\C\mapsto\C^{n_y}$, so that the closed loop becomes
\begin{align}\label{eq:CL}
y(s)=S(s)d(s)+T(s)r(s)-T(s)n(s).
\end{align}
Below the closed-loop bandwidth, it holds that $\twonorm{r(\jw)}\gg\twonorm{S(\jw)d(\jw)}$ and $\twonorm{r(\jw)}\gg\twonorm{n(\jw)}$, allowing $T(s)$ to be estimated from~\eqref{eq:CL}, even for small $\twonorm{r(s)}$. However, due to the large condition number of $R$, setting $r(s)=e_i \rho_i(s)$ may lead to large actuator gains or require to limit the amplitude of $r(s)$, impacting the accuracy of the estimates, $\hat{T}(s)$ and $\hat{S}(s)$. To address this, the modal transformation is applied to~\eqref{eq:CDsystem} and the reference signal designed in modal space, tuning $r(s)$ to the gain and bandwidth of each mode. A non-parametric estimate of $\hat{T}(\jw)$ is then obtained in mode space using SISO system identification techniques~\cite{LJUNG}.

Alternatively, one could consider parametric methods for estimating sensitivity in closed loop~\cite{chevalier2015closed, de2017minimal, guzman2014itcli}. While these methods are applicable to a broader range of systems than~\eqref{eq:CDsystem}, they require modeling and parameter identification in a high-dimensional space, making their implementation on large-scale systems such as~\eqref{eq:CDmodal} challenging. This challenge is exacerbated by the large condition number of $R$, which can lead to numerical instabilities if the structure of~\eqref{eq:CDmodal} is not explicitly considered. In contrast, the method proposed here explicitly considers the structure of~\eqref{eq:CDsystem}.

This paper is organised as follows. Section~\ref{sec:modal} summarises the modal transformation. In Section~\ref{sec:reference}, the reference signal is designed that is used in Section~\ref{sec:id} to estimate the sensitivity mode-by-mode. Finally, the approach is applied to Diamond's electron beam stabilisation problem in Section~\ref{sec:simulations}.

\emph{Notation and Definitions} For a scalar, vector or matrix $A$, let $\trans{A}$ ($\herm{A}$) be its (Hermitian) transpose, $\diag(A_1,\dots,A_n)$ a diagonal matrix with diagonal elements $A_1,\dots,A_n$. Let $I_n$ denote the identity matrix in $\R^{n\times n}$. For a matrix $A$, let $\pinv{A}$ denote the pseudo-inverse~\cite[p. 290]{GOLUB4}, $\twonorm{A}$ the spectral norm, and $\kappa(A)\eqdef\twonorm{A}/\twonorm{\pinv{A}}$ the condition number.

\section{BACKGROUND: MODAL REPRESENTATION\label{sec:modal}}

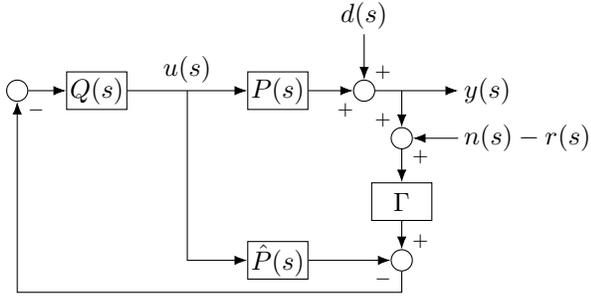
\begin{figure}
\centering
\begin{tikzpicture}[align=center,node distance=1cm,
	    every node/.style={inner sep=2pt,rectangle, minimum height=1em, text centered},
	    block/.style={draw,inner sep=1pt,rectangle, minimum width=0.6cm, minimum height=0.5cm},
	    sum/.style={draw,minimum height = 8pt,circle,inner sep=1pt},
	    nnode/.style={draw,fill=black,minimum height=2pt,circle,inner sep=0pt}]
\node[sum] (s1) {};
\node[block, right = 0.65cm of s1.center] (c) {$Q(s)$};
\node[block, right = 2cm of c.center] (g) {${P(s)}$};
\node[block, below = 2.75cm of g.center] (gm) {$\hat{P}(s)$};
\node[sum,   right = 1cm of g.center] (s3) {};
\node[right = 2cm of s3.center] (y) {$y(s)$};
\node[above = 1cm of s3.center] (dist) {$d(s)$};
\node[below = 3.5cm of s1.center] (fb) {};
\node[anchor=center] (tmp) at ($(s3.center)!0.5!(y.west)$) {};
\node[sum, anchor=center] (s4) at (gm.center -| tmp.center){};
\node[block] (gamma) at ($(s4.north)!0.2!(tmp.center)$) {$\Gamma$};
\node[sum, anchor=center] (s5) at ($(s4.north)!0.8!(tmp.center)$) {};
\node[sum, anchor=center] (s6) at ($(s4.north)!0.5!(tmp.center)$) {};
\node[inner sep=0pt] (sec) at ($(c)!0.5!(g)$) {};
\node[anchor=west] (n) at (s5.center -| y.west){$n(s)$};
\node[anchor=west] (r) at (s6.center -| y.west){$r(s)$};

\draw[-Latex] (s1.east) -- (c.west);
\draw[-Latex] (c.east) -- (g.west);
\draw[-Latex] (g.east) -- (s3.west);
\draw[-Latex] (tmp.center) -- (s5.north);
\draw[-Latex] (s5.south) -- (s6.north);
\draw[-Latex] (s6.south) -- (gamma.north);
\draw[-Latex] (gamma.south) -- (s4.north);
\draw[-Latex] (s3.east) -- (y.west);
\draw[-Latex] (gm.east) -- (s4.west);
\draw[-Latex] (sec.center) |- (gm.west);
\draw[-Latex] (s4.south) |- (fb.center) -| (s1.south);
\draw[-Latex] (dist.south) -| (s3.north);
\draw[-Latex] (n.west) -- (s5.east);
\draw[-Latex] (r.west) -- (s6.east);
\node[] at ([yshift=0.1cm]sec.north) {$u(s)$};
\node[font=\scriptsize] at ([xshift=0.25cm,yshift=0.25cm]s4.center) {$+$};
\node[font=\scriptsize] at ([xshift=-0.25cm,yshift=-0.25cm]s4.center) {$-$};
\node[font=\scriptsize] at ([xshift=0.25cm,yshift=-0.25cm]s1.center) {$-$};
\node[font=\scriptsize] at ([xshift=0.25cm,yshift=0.25cm]s3.center) {$+$};
\node[font=\scriptsize] at ([xshift=-0.25cm,yshift=-0.25cm]s3.center) {$+$};
\node[font=\scriptsize] at ([xshift=-0.25cm,yshift=0.25cm]s5.center) {$+$};
\node[font=\scriptsize] at ([xshift=0.25cm,yshift=-0.25cm]s5.center) {$+$};
\node[font=\scriptsize] at ([xshift=-0.25cm,yshift=0.25cm]s6.center) {$+$};
\node[font=\scriptsize] at ([xshift=0.25cm,yshift=-0.25cm]s6.center) {$-$};

\node (tmp0) at ([xshift=-0.25cm]s1.west |- dist.south) {};
\node (tmp1) at ([xshift=-0.3cm]g.west |- dist.south) {};
\node (tmp4) at ([xshift=0.5cm]$(s5.center)!0.5!(s6.center)$) {};
\node (tmp3) at ([xshift=-0.3cm]g.west |- tmp4.center) {};
\node (tmp5) at ([yshift=-0.25cm]fb.center -| tmp4.center) {};
\node (tmp6) at ([xshift=-0.25cm,yshift=-0.25cm]s1.west |- fb.center) {};

\draw[densely dotted] (tmp0.center) -- (tmp1.center) -- (tmp3.center) -- (tmp4.center) -- (tmp5.center) -- (tmp6.center) -- (tmp0.center);

\node[inner sep=5pt, anchor=north west] () at (tmp0.center) {$C(s)$};
\end{tikzpicture}
\caption{Controller structure with plant $P(s)$, plant model $\hat{P}(s)$, IMC filter $Q(s)$, compensator $\Gamma$, disturbance $d(s)$, noise $n(s)$, and reference signal $r(s)$.\label{fig:diagram}}
\end{figure}

Although our method is applicable to any controller structure for CD systems, this paper focuses on the IMC structure used at Diamond~\cite{SANDIRACONTROLDESIGN}, as shown in Fig.~\ref{fig:diagram}. To design the reference $r(s)$, the MIMO representation~\eqref{eq:CDsystem} is mapped to modal space by substituting the thin SVD, $R=U\Sigma\trans{V}$~\cite{HEATH}:
\begin{align}\label{eq:CDmodal}
\tilde{y}(s)=\Sigma g(s)\tilde{u}(s) + \tilde{d}(s),
\end{align}
where $\tilde{y}(s)\eqdef\trans{U}y(s)$, $\tilde{d}(s)\eqdef\trans{U}d(s)$, and $\tilde{u}(s)\eqdef\trans{V}u(s)$. The matrices $U$ and $V$ satisfy $\xTx{U}=I$ and $\xTx{V}=I$ and $\Sigma\eqdef\diag(\sigma_1,\dots,\sigma_{n_y})$ is a diagonal matrix containing the singular values. Throughout the paper it is assumed that $\rank(R)=n_y\leq n_u$, which holds for Diamond, but our results remain valid for other configurations.

In modal space, the IMC filter $\tilde{Q}(s)\eqdef \trans{V} Q(s) U$ is diagonal with elements~\cite{SANDIRACONTROLDESIGN} 
\begin{align}
q_i(s)\eqdef\frac{\lambda(s)}{\sigma_i g(s)},
\end{align} where $\lambda(s)$ contains the non-minimum phase parts of $g(s)$ and shapes the overall bandwidth. The compensator $\tilde{\Gamma}\eqdef \trans{U} \Gamma U$ attenuates controller gains for small $\sigma_i$ and is diagonal with elements $\gamma_i\eqdef\sigma_i^2/(\sigma_i^2+\mu)$, where $\mu >0$ is a scalar regularisation parameter. For an accurante plant model ($\hat{P}(s)\equiv P(s)$), this results in the modal inputs (see~\cite{PHDIDRIS})
\begin{align}\label{eq:inputmodebymode}
\tilde{u}_i(s)=-\frac{\gamma_i}{\sigma_i}\frac{\lambda(s)/g(s)}{1-(1-\gamma_i)\lambda(s)}(\tilde{d}_i(s)+\tilde{n}_i(s)-\tilde{r}_i(s)),
\end{align}
and the modal outputs
\begin{align}\label{eq:CLmodebymode}
\tilde{y}_i(s)=\tilde{S}_i(s)\tilde{d}_i(s)+\tilde{T}_i(s)(\tilde{r}_i(s)-\tilde{n}_i(s)), 
\end{align}
for $i=1,\dots,n_y$, and where $\tilde{S}_i(s)=1-\tilde{T}_i(s)$ and
\begin{align}\label{eq:Tmodebymode}
\tilde{T}_i(s)\reqdef \gamma_i\frac{\lambda(s)}{1-(1-\gamma_i)\lambda(s)}.
\end{align}

The sensitivities in original space are obtained as 
\begin{align*}
&T(s)\eqdef U\diag(\tilde{T}_1(s),\dots,\tilde{T}_{n_y}(s))\trans{U},
&S(s)\eqdef U\diag(\tilde{S}_1(s),\dots,\tilde{S}_{n_y}(s))\trans{U}. 
\end{align*}
The minimum and maximum gains of $S(s)$ are shown in Fig.~\ref{fig:modebymodeT} for the Diamond system (see Section~\ref{sec:simulations}), where $\twonorm{S(\jw)}\equiv\abs{\tilde{S}_{n_y}(\jw)}$ and $1/\twonorm{S^{-1}(\jw)}\equiv 1/\abs{\tilde{S}_1(\jw)}$ for frequencies below \SI{100}{\Hz}. Due to the large $\kappa(R)$, the compensator $\Gamma$ effectively reduces the bandwidth for higher-order modes, leading to a significant difference between minimum and maximum gains of $S(s)$. 

\begin{figure}[]

    \centering
    
    \begin{subfigure}[t]{0.4\columnwidth}%
    \centering
        \begin{tikzpicture}
        \begin{axis}[view={0}{90},
        width=0.9\linewidth, height=0.9\linewidth,
        xmode=log, point meta min=-2, point meta max=1,
        ylabel={Mode $i$},ytick align=outside,ytick={1,30,60,90,120,150},
        xlabel={Frequency (\si{\Hz})},xtick align=outside,tick pos=left,
        colormap/viridis,
        colorbar horizontal,
        colorbar style={
        at={(0,0.9\linewidth-1.1cm)},anchor=north west, /pgf/number format/precision=0, /pgf/number format/fixed, /pgf/number format/fixed zerofill,
        xtick align=outside,tick pos=right,xticklabel pos=top, xmin=-2, xmax=1, xtick = {-5,-4,-3,-2,-1,0,1,2},xlabel={Magnitude (\si{\micro\meter\per\sqrt{\Hz}})}, xlabel style={yshift=-0.4em},height=0.75em,
        }, shader=flat]
        \addplot3[surf,mesh/rows=37,shader=interp,draw=none] table [x=freq, y=mode, z expr=log10(sqrt(10^\thisrow{psd_off})), col sep=comma] {disturbance_psd.csv};
        \end{axis}
        \end{tikzpicture}
    \caption{Open-loop: ASDs of $\tilde{y}_i^\mathrm{ol}(s)$.}\label{fig:fofb_off}
    \end{subfigure}\hfill
    \centering
    \begin{subfigure}[t]{0.4\columnwidth}%
    \centering
        \begin{tikzpicture}
        \begin{axis}[view={0}{90},
        width=0.9\linewidth, height=0.9\linewidth,
        xmode=log,point meta min=-2, point meta max=1,
        ylabel={Mode $i$},ytick align=outside,ytick={1,30,60,90,120,150},
        xlabel={Frequency (\si{\Hz})},xtick align=outside,tick pos=left,
        colormap/viridis,
        colorbar horizontal,
        colorbar style={
        at={(0,0.9\linewidth-1.1cm)},anchor=north west, /pgf/number format/precision=0, /pgf/number format/fixed, /pgf/number format/fixed zerofill,
        xtick align=outside,tick pos=right,xticklabel pos=top, xmin=-2, xmax=1, xtick ={-5,-4,-3,-2,-1,0,1,2},
        xlabel={Magnitude (\si{\micro\meter\per\sqrt{\Hz}})}, xlabel style={yshift=-0.4em},height=0.75em,
        }, shader=flat]
        \addplot3[surf,mesh/rows=37,shader=interp,draw=none] table [x=freq, y=mode, z expr=log10(sqrt(10^\thisrow{psd_on})), col sep=comma] {disturbance_psd.csv};
        \addplot3 [mark=none, color=white, thick, dotted] table [x=bw_S, y=mode, col sep=comma] {bandwidths.csv};
        \end{axis}
        \end{tikzpicture}
    \caption{Closed-loop: ASDs of $\tilde{y}_i^\mathrm{cl}(s)$ and bandwidths of $\tilde{S}_i(s)$ (dotted).}\label{fig:fofb_on}
    \end{subfigure}\\[0.5em] 
\caption{Spectral density of output in modal space.}\label{fig:fofb}
\end{figure}
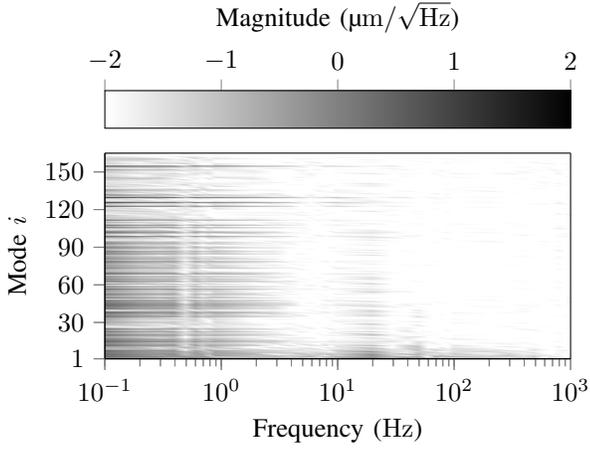
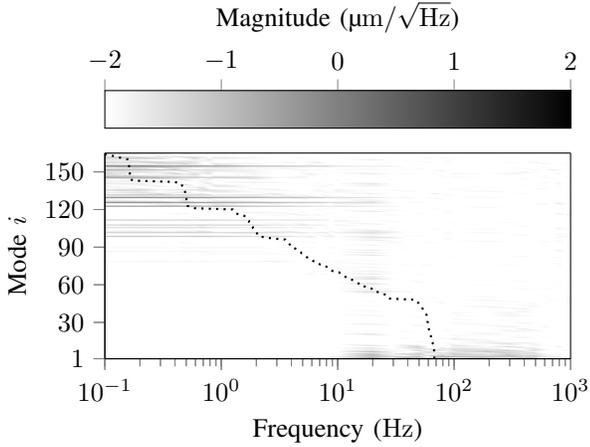

The reduction in bandwidth for higher-order modes is justified by the characteristic spectrum of $\tilde{d}(s)$. The amplitude spectral density (ASD) of the output in mode space for disabled FOFB, i.e.\ when $\tilde{y}^\mathrm{ol}(s)=\tilde{d}(s)+\tilde{n}(s)$, is shown in Fig.~\ref{fig:fofb_off}. For low frequencies for which $\twonorm{\tilde{d}(\jw)}\gg\twonorm{\tilde{n}(\jw)}$, the spectrum of $\tilde{y}^\mathrm{ol}(s)$ is proportional to the square of the singular values~\cite{kempf:ibic2023-mo2i02}. The simulated attenuation in closed loop is shown in Fig.~\ref{fig:fofb_on}, i.e.\ when $\tilde{y}^\mathrm{cl}(s)=\tilde{S}(s)\tilde{d}(s)-\tilde{T}(s)\tilde{n}(s)$. The dashed line represents the bandwidth (\SI{-3}{\dB} frequency) of $\tilde{S}_i(s)$.

\begin{figure}

\centering

\begin{tikzpicture}
\begin{semilogxaxis}[width=0.8\linewidth, height=0.4\linewidth,
xmin=0.01, xmax=5000, ymin=-10,ymax=5, minor ytick={-20,-15,...,30},
ylabel={Magnitude (\si{\dB})}, xlabel={Frequency (\si{\hertz})},
grid=both, major grid style={line width=0.5pt,draw=black},
]
\addplot[draw=none,forget plot,name path=Smin1]   table [x=w, y=miny, col sep=comma] {S_imc9.csv};
\addplot[draw=none,forget plot,name path=Smax1]  table [x=w, y=maxy, col sep=comma] {S_imc9.csv};
\addplot[legbluefill,forget plot,opacity=0.2] fill between[of=Smin1 and Smax1];
\addplot[black,very thick]   table [x=w, y=miny, col sep=comma] {S_imc9.csv};
\addplot[black,very thick]  table [x=w, y=maxy, col sep=comma] {S_imc9.csv};
\node[anchor=center,align=center,font=\normalsize,fill=white,opacity=1,draw=none,inner sep=1pt] (Sm) at (1,0) {$\twonorm{S(\jw)}$};
\node[anchor=center,align=center,font=\normalsize,fill=white,opacity=1,draw=none,inner sep=1pt] (Sm) at (40,-7.5) {$1/\twonorm{\inv{S}(\jw)}$};
\end{semilogxaxis}
\end{tikzpicture}
\caption{Minimum and maximum sensitivity gains.} \label{fig:modebymodeT}
\end{figure}
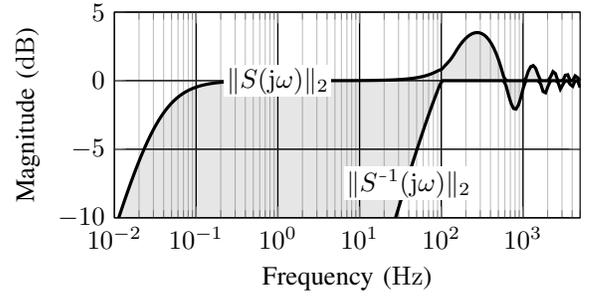

\section{REFERENCE SIGNAL DESIGN\label{sec:reference}}

According to~\eqref{eq:inputmodebymode}, the output reference signal $r(s)=e_i\rho_i(s)$ produces large control inputs if $e_i$ is aligned with a column $U_i$ of $U$ corresponding to a higher-order mode with a small singular value $\sigma_i$. In contrast, fixing the reference in modal space as  
\begin{align}\label{eq:rhomodebymode}
&\tilde{r}(s)=e_i\rho_i(s),
\end{align}
so that $r(s)=U_i\rho_i(s)$, allows~\eqref{eq:inputmodebymode} and~\eqref{eq:CLmodebymode} to be tuned mode-by-mode through adapting $\rho_i(s)$. Moreover, with the reference~\eqref{eq:rhomodebymode} applied to mode $i$, it holds that, 
\begin{align}
\tilde{y}_j(s)=
\begin{cases}
\Delta\tilde{y}_j(s)+\tilde{T}_j(s)\rho_j(s) & \text{for }j=i,\\
\Delta\tilde{y}_j(s) & \text{otherwise},\\
\end{cases}
\end{align}
where
\begin{align}\label{eq:deltay}
\Delta\tilde{y}_i(s)\eqdef \tilde{S}_i(s)\tilde{d}_i(s) -\tilde{T}_i(s)\tilde{n}_i(s),\quad i=1,\dots,n_y.
\end{align}
The (complementary) sensitivity can therefore be estimated from $\rho_i(s)$ and $\tilde{y}_i(s)$ using SISO techniques.

\subsection{Lower Bound}

The terms~\eqref{eq:deltay} introduce an estimation error. For an estimate $\hat{\tilde{T}}_i(\jw)\eqdef\tilde{y}_i(\jw)/\rho_i(\jw)$, the absolute estimation error is 
\begin{align}\label{eq:error}
\tilde{\epsilon}_i(\jw)\eqdef \abs{\hat{\tilde{T}}_i(\jw)-\tilde{T}_i(\jw)}=\abs{\Delta\tilde{y}_i(\jw)/\rho_i(\jw)}. 
\end{align}

To bound $\tilde{\epsilon}_i(\jw)\leq\epsilon_\text{max}$ for frequencies of interest, $\omega\leq\hat{\omega}_i$, the reference signal must therefore satisfy the lower bound
\begin{align}\label{eq:minchirp}
\abs{\rho_i(\jw)}\geq\frac{1}{\epsilon_\text{max}}\times\abs{\tilde{S}_i(\jw)\tilde{d}_i(\jw) -\tilde{T}_i(\jw)\tilde{n}_i(\jw)},
\end{align}
Since $\xTx{U}=I$, the resulting estimation error in original space is then bounded by
\begin{align}\label{eq:errorboundT}
\twonorm{T(\jw)-\hat{T}(\jw)}=\twonorm{\tilde{T}(\jw)-\hat{\tilde{T}}(\jw)}\leq\epsilon_\text{max}.
\end{align}

\subsection{Upper Bound}

Although a large $\abs{\rho_i(\jw)}$ reduces the estimation error~\eqref{eq:error}, system~\eqref{eq:CDsystem} is limited by component-wise input and output constraints, $\abs{u_i(t)}\leq u_\text{max}$ and $\abs{y_i(t)}\leq y_\text{max}$~\cite{EBSCONF}. Assuming that $r(s)$ is a cosine at frequency $\omega$ with $\abs{r(\jw)}\gg\max(\abs{d(\jw)},\abs{n(\jw)})$, the input and output constraints are approximated in frequency domain by
\begin{align}\label{eq:freqdomainconstraints}
&\abs{u_i(\jw)}\leq u_\text{max},
&\abs{y_i(\jw)}\leq y_\text{max}.
\end{align}
Since $\xTx{V}=I$, it holds that $\abs{u_i(\jw)}\leq\twonorm{u(\jw)}=\twonorm{\tilde{u}(\jw)}$, and assuming that $\abs{\tilde{r}_i(\jw)}\gg \abs{\tilde{d}_i(\jw)-\tilde{n}_i(\jw)}$ for $\omega\leq\hat{\omega}_i$ in~\eqref{eq:inputmodebymode}, $\twonorm{\tilde{u}(\jw)}\approx\abs{\tilde{u}_i(\jw)}$ for a reference as in~\eqref{eq:rhomodebymode}. To limit $\abs{u_i(\jw)}\leq u_\text{max}$, the reference must therefore satisfy
\begin{align}\label{eq:maxchirp}
\abs{\rho_i(\jw)}\leq \frac{u_\text{max}\sigma_i}{\gamma_i}\times\frac{1-(1-\gamma_i)\lambda(\jw)}{\lambda(\jw)/g(\jw)}.
\end{align}
Similarly, to limit $\abs{y_i(\jw)}\leq y_\text{max}$, the reference must satisfy
\begin{align}\label{eq:maxchirp2}
\abs{\rho_i(\jw)}\leq y_\text{max}.
\end{align}

\begin{figure}
\centering
\begin{tikzpicture}
\begin{semilogyaxis}[width=0.8\linewidth, height=0.4\linewidth,
xmin=1, xmax=165, 
ylabel={Magnitude (\si{\micro\meter})}, xlabel={Mode $i$ (-)},
grid=both, major grid style={line width=0.5pt,draw=black},
]
\addplot[black,very thick, dashed]   table [x=mode, y=lb, col sep=comma] {lb_ub.csv};
\addplot[black,very thick]  table [x=mode, y=ub, col sep=comma] {lb_ub.csv};
\end{semilogyaxis}
\end{tikzpicture}
\caption{Lower (dashed) and upper bounds on the reference amplitude for $u_\text{max}=\SI{1}{\A}$, $y_\text{max}=\SI{150}{\micro\meter}$, and $\epsilon_\text{max}=0.1$.}\label{fig:bounds}
\end{figure}
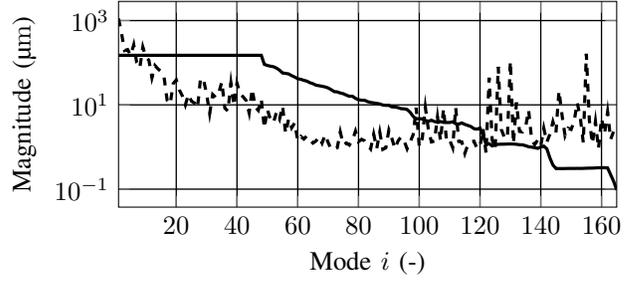

The lower and upper bounds for the case that $\rho_i(\jw)$ is a chirp signal (Section~\ref{sec:id}) with amplitude $A_i$ are shown in Fig.~\ref{fig:bounds}. The upper bound (solid) is computed from the minimum of~\eqref{eq:maxchirp} and~\eqref{eq:maxchirp2}, and the lower bound (dashed) from~\eqref{eq:minchirp} for $\epsilon_\text{max}=0.1$ and includes an additional factor introduced by windowing (Section~\ref{sec:id}). Due to the large $\kappa(R)$, higher-order modes ($i\geq 100$) require significantly larger control inputs than lower-order modes. However, the upper bound~\eqref{eq:maxchirp} limits $A_i$, which can result in estimation errors larger than $\epsilon_\text{max}$. The large $\kappa(R)$ also impacts lower-order modes through large disturbances (Fig.~\ref{fig:fofb_off}), resulting in a low signal-to-noise ratio. Even though~\eqref{eq:maxchirp} would allow for larger $A_i$, lower-order modes are limited by~\eqref{eq:maxchirp2}.

\section{SENSITIVITY IDENTIFICATION\label{sec:id}}

To identify the sensitivity, the reference signal is swept from $\omega=0$ to $\omega=\hat{\omega}_i$ for each mode:\ 
\begin{align*}
&\rho_i(t)=A_i\cos(\frac{\hat{\omega}_i t}{N T_s} t),
&t\in [0,N T_s],
&&i=1,\dots,n_y,
\end{align*}
where $N$ is the number of samples and $T_s$ the sample time. To avoid input and output saturation, the maximum frequency $\hat{\omega}_i$ of the reference signal for each mode is set to $5$ times the bandwidth of~\eqref{eq:Tmodebymode}. The amplitude $A_i$ is set to the upper bound from Fig.~\ref{fig:bounds}. 

After mapping the closed-loop time series data to mode space, the estimate $\hat{\tilde{T}}_i(\jw)$ is obtained from the Blackman-Tuckey spectral analysis method \cite[Ch.\ 6]{LJUNG}:
\begin{align}\label{eq:tfest}
\hat{\tilde{T}}_i(\jw) \eqdef \frac{\displaystyle\hat{{\Phi}}_{\tilde{y}_i{\rho}_i}(\omega)}{\displaystyle\hat{\Phi}_{{\rho}_i{\rho}_i}(\omega)} \eqdef \frac{\displaystyle\sum^{M}_{\tau = -M}\hat{R}_{\tilde{y}_i{\rho}_i}(\tau)W_{M}(\tau)e^{\jw t}}{\displaystyle\sum^{M}_{\tau = -M}\hat{R}_{{\rho}_i{\rho}_i}(\tau)W_{M}(\tau)e^{\jw t}},
\end{align}
where $W_{M}(\tau)$ is a Hamming window and the correlation functions are computed in modal space as
\begin{align}
\hat{R}_{v_i w_i}(\tau) &\eqdef \frac{1}{N} \sum^{N}_{t = 1}v_i(t + \tau)w_i(t).
\end{align}
The windowing function allows for a smoothed spectral estimate of the complementary sensitivity by computing a weighted average of the frequency response of neighboring points. Large windows (small $M$) filter out the variance from~\eqref{eq:error} but introduce bias, which is particularly visible for low frequencies at which $\tilde{T}_i(s)\approx 1$. The converse is true for small windows, and the trade-off between bias and variance must be considered when selecting an $M$ value (note that $M$ is denoted as $\gamma$ in \cite[Ch.\ 6]{LJUNG}). Here, the parameter $M$ is chosen proportionally to $1/\hat{\omega}_i$, ranging from $M=500$ for mode $i=1$ to $M=14000$ for $i=n_y$. Note that a windowing factor is included in the lower bound of Fig.~\ref{fig:bounds}.

Although the complementary sensitivity is estimated in closed-loop, the reference $r(s)=e_i\rho(s)$ is not correlated with the disturbance $d(s)$ or the noise $n(s)$, avoiding closed-loop issues encountered in plant identification~\cite{VANDENHOF1998173}. The identification procedure is summarised in Algorithm~\ref{alg:sysid}, where $U_i$ refers to column $i$ of the modal transformation matrix $U$ (Section~\ref{sec:modal}). Neglecting the complexity of the data collection and the mapping of the signals to mode space, Algorithm~\ref{alg:sysid} is of complexity $n_y\times\mathcal{O}(\hat{\tilde{T}}_i)$, where $\mathcal{O}(\hat{\tilde{T}}_i)$ is the complexity of estimating the scalar transfer function $\hat{\tilde{T}}_i$. Without the modal representation, estimating $T(s)$ would be of complexity $n_y^2\times\mathcal{O}(\hat{\tilde{T}}_i)$.

\begin{algorithm}
\caption{Sensitivity identification in modal space.}\label{alg:sysid}
\begin{algorithmic}[1]
\INPUT $y_\text{max}$, $u_\text{max}$, $\epsilon_\text{max}$
\OUTPUT $S(\jw)$
\FOR{$i = 1$ to $n_y$}
    \STATE Compute $\rho_i(t)$ according to~\eqref{eq:minchirp} and~\eqref{eq:maxchirp}--\eqref{eq:maxchirp2}
	\STATE Collect closed-loop data $y^\mathrm{cl}(t)$ for $r(t)=U_i \rho_i(t)$
	\STATE Map to modal space via $\tilde{y}^\mathrm{cl}(t)=\trans{U}y^\mathrm{cl}(t)$
	\STATE Compute $\hat{\tilde{T}}_i(\jw)$ using $\tilde{y}^\mathrm{cl}(t)$ and $\rho_i(t)$
\ENDFOR
\STATE{Set $\hat{T}(\jw)=U\hat{\tilde{T}}_i(\jw)\trans{U}$ and $S(\jw)=I-T(\jw)$}
\end{algorithmic}
\end{algorithm}

\section{CASE STUDY: DIAMOND LIGHT SOURCE\label{sec:simulations}}

At Diamond, the FOFB uses $n_y=173$ sensors and $n_u=172$ magnets operated at $f_s=\SI{10}{\kHz}$. However, the FOFB can be reconfigured, allowing any combination of $n_y\leq 173$ sensors and $n_u\leq 172$ outputs and inputs. For these simulations, it is assumed that $n_y=n_u=165$ with $\kappa(R)=9837$ ($\sigma_\text{max}=195$ and $\sigma_\text{min}=0.02$), $u_{max}=\SI{5}{\A}$, and $y_{max}=\SI{150}{\micro\meter}$. The actuator dynamics are $g(s)=a/(s+a)\mathrm{e}^{-\tau_d s}$ with $a=2\pi\times 700 \si{\radian\per\second}$ and a time delay $\tau_d=\SI{900}{\micro\second}$~\cite{PHDIDRIS}. The transfer function $\lambda(s)$ is $\lambda(s)=\bar{\lambda}/(s+\bar{\lambda})\mathrm{e}^{-\tau_d s}$ with $\bar{\lambda}=2\pi\times 176 \si{\radian\per\second}$ and the regularisation parameter is set to $\mu=1$. As reflected in Fig.~\ref{fig:modebymodeT}, the large time delay causes a sensitivity overshoot of \SI{3.5}{\dB} and the large $\kappa(R)$ bandwidths of $\tilde{S}_i(s)$ that spread from \SIrange{0.03}{70}{\Hz} (Fig.~\ref{fig:fofb_off}).

To evaluate the Algorithm~\ref{alg:sysid} and verify the bound~\eqref{eq:minchirp} with $\epsilon_\text{max}=0.1$ for Diamond, the closed-loop system~\eqref{eq:CL} was simulated using $10^5$ measured disturbance samples from Diamond, which were different from the $10^4$ samples used to compute the bounds in Fig.~\ref{fig:bounds}. The reference signal was chosen as in Section~\ref{sec:id} and Algorithm~\ref{alg:sysid} evaluated $10$ times for $N=10^4$ samples. The estimates were computed using Matlab System Identification Toolbox on a desktop computer (Intel i7-7700 CPU @ \SI{3.1}{\giga\Hz}, \SI{8}{\giga\byte}) within less than \SI{3}{\minute} for all 165 modes. 

\begin{figure}[]
    \centering
    \begin{subfigure}[t]{0.33\columnwidth}%
    \centering
        \begin{tikzpicture}
        \begin{axis}[view={0}{90},
        width=\linewidth, height=\linewidth,
        xmode=log, point meta min=-10, point meta max=5,
        ylabel={Mode $i$},ytick align=outside,ytick={1,30,60,90,120,150},
        xlabel={Normalised frequency (-)},xtick align=outside,tick pos=left,
        colormap/viridis,
        colorbar horizontal,
        colorbar style={
        at={(0,\linewidth-1.1cm)},anchor=north west, /pgf/number format/precision=0, /pgf/number format/fixed, /pgf/number format/fixed zerofill,
        xtick align=outside,tick pos=right,xticklabel pos=top, xmin=-10, xmax=5, xtick = {-10,-5,0,5},
        xlabel={Magnitude (\si{\dB})},xlabel style={yshift=-0.4em},height=0.75em,
        }, shader=flat]
        \addplot3[surf,mesh/rows=28,shader=interp,draw=none] table [x=freq, y=mode, z=T, col sep=comma] {colorplots.csv};
        \end{axis}
        \end{tikzpicture}
    \caption{Magnitude of ${\tilde{T}_i(\jw)}$.}\label{fig:trueT}
    \end{subfigure}
    \centering
    \begin{subfigure}[t]{0.33\columnwidth}%
    \centering
        \begin{tikzpicture}
        \begin{axis}[view={0}{90},
        width=\linewidth, height=\linewidth,
        xmode=log, point meta min=-10, point meta max=5,
        ylabel={Mode $i$},ytick align=outside,ytick={1,30,60,90,120,150},
        xlabel={Normalised frequency (-)},xtick align=outside,tick pos=left,
        colormap/viridis,
        colorbar horizontal,
        colorbar style={
        at={(0,\linewidth-1.1cm)},anchor=north west, /pgf/number format/precision=0, /pgf/number format/fixed, /pgf/number format/fixed zerofill,
        xtick align=outside,tick pos=right,xticklabel pos=top, xmin=-10, xmax=5, xtick = {-10,-5,0,5},
        xlabel={Magnitude (\si{\dB})},xlabel style={yshift=-0.4em},height=0.75em,
        }, shader=flat]
        \addplot3[surf,mesh/rows=28,shader=interp,draw=none] table [x=freq, y=mode, z=Test, col sep=comma] {colorplots.csv};
        \end{axis}
        \end{tikzpicture}
    \caption{Magnitude of ${\hat{\tilde{T}}_i(\jw)}$.}\label{fig:estimT}
    \end{subfigure}
    \centering
    \begin{subfigure}[t]{0.33\columnwidth}%
    \centering
        \begin{tikzpicture}
        \begin{axis}[view={0}{90},
        width=\linewidth, height=\linewidth,
        xmode=log, point meta min=-1, point meta max=1,
        ylabel={Mode $i$},ytick align=outside,ytick={1,30,60,90,120,150},
        xlabel={Normalised frequency (-)},xtick align=outside,tick pos=left,
        colormap/viridis,
        colorbar horizontal,
        colorbar style={
        at={(0,\linewidth-1.1cm)},anchor=north west, /pgf/number format/precision=2, /pgf/number format/fixed, /pgf/number format/fixed zerofill,
        xtick align=outside,tick pos=right,xticklabel pos=top, xmin=-1, xmax=1, xtick = {-1, -0.5, 0, 0.5, 1},
        xticklabels= {$10^{-1}$, $10^{-0.5}$, $10^{0}$, $10^{0.5}$, $10^{1}$}, xlabel={Magnitude (-)},xlabel style={yshift=-0.4em},height=0.75em,
        }, shader=flat]
        \addplot3[surf,mesh/rows=28,shader=interp,draw=none] table [x=freq, y=mode, z=Terr, col sep=comma] {colorplots.csv};
        \end{axis}
        \end{tikzpicture}
    \caption{Absolute error, $\abs{\tilde{T}_i(\jw)-\hat{\tilde{T}}_i(\jw)}$.}\label{fig:estimerror}
    \end{subfigure}\\[0.5em]
\caption{Complementary sensitivities and estimation error over modes and frequencies. For each mode, the normalised frequency ranges from 0 to $\tilde{\omega}_i$.}\label{fig:estimall}
\end{figure}
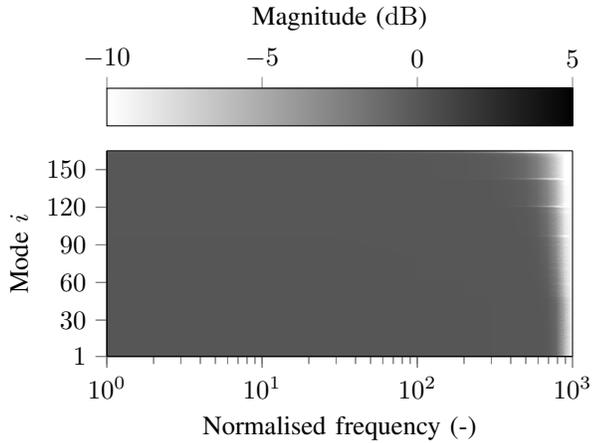
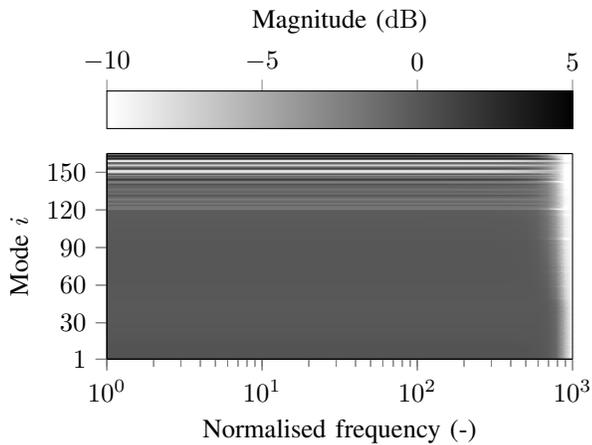
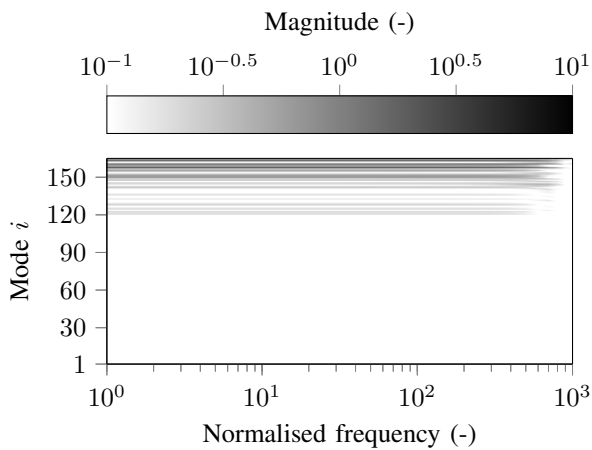

The magnitudes of the true complementary sensitivity in modal space, $\tilde{T}(\jw)$, the average of the estimates, $\mathrm{E}\lbrace\hat{\tilde{T}}(\jw)\rbrace$, and the mean absolute error, $\mathrm{E}\lbrace\abs{\hat{\tilde{T}}(\jw)-\tilde{T}(\jw)}\rbrace$ are shown in Fig~\ref{fig:trueT}--\ref{fig:estimerror}. The horizontal axis refers to the normalised frequency that ranges from 0 to $\hat{\omega}_i$ for each mode. Fig.~\ref{fig:estimerror} shows that for higher-order modes for which the lower bounds from Fig.~\ref{fig:bounds} are violated, the resulting estimation error is larger than $\epsilon_\text{max}$. However, for lower-order modes, the estimation error is below $\epsilon_\text{max}$ as expected from Fig.~\ref{fig:bounds}. 

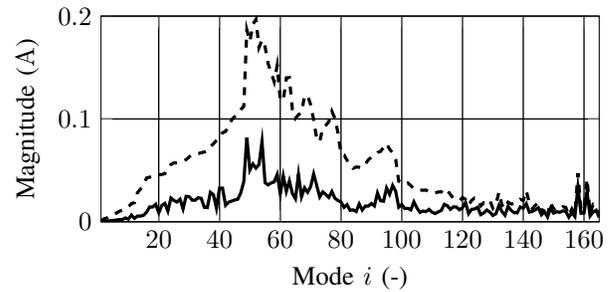
\begin{figure}
\centering
\begin{tikzpicture}
\begin{axis}[width=0.8\linewidth, height=0.4\linewidth,
xmin=1, xmax=165, ymin=0,ymax=0.2,
ylabel={Magnitude (\si{\A})}, xlabel={Mode $i$ (-)},ytick={0,0.1,0.2},
grid=both, major grid style={line width=0.5pt,draw=black},
]
\addplot[black,very thick, dashed]   table [x=mode, y=mode_max, col sep=comma] {input_limits.csv};
\addplot[black,very thick]  table [x=mode, y=orig_max, col sep=comma] {input_limits.csv};
\end{axis}
\end{tikzpicture}
\caption{Maximum input amplitudes in mode space (dashed) and original space (solid) for each iteration of Algorithm~\ref{alg:sysid}.}\label{fig:inputs}
\end{figure}

The maximum input amplitudes $\max_j\abs{\tilde{u}_j(t)}$ and $\max_j\abs{u_j(t)}$ for a reference signal on mode $i$ are shown in Fig.~\ref{fig:inputs}. For all modes, it holds that $\max_i\abs{\tilde{u}_i(\jw)}\leq \SI{1.5}{\A}$, which is below the limit $u_\text{max}=\SI{5}{\A}$. This is related to the frequency-domain approximation~\eqref{eq:freqdomainconstraints} and to choosing a constant (frequency-independent) amplitude of the chirp. As expected from the orthogonality from the modal transformation, it holds that $\max_j\abs{\tilde{u}_j(t)}\leq\max_j\abs{u_j(t)}$.

\section{CONCLUSION\label{sec:conclusion}}

In this paper, we have proposed an algorithm for closed-loop sensitivity identification for ill-conditioned cross-directional systems and evaluated it using Diamond's electron beam stabilisation problem. While the controller was fixed to a standard structure used in electron beam stabilisation, an additional output reference signal was introduced in closed loop. By aligning the reference signal with each mode, the MIMO identification problem was reduced to a SISO identification problem, allowing the sensitivity to be estimated mode-by-mode using SISO techniques. We derived lower and upper bounds on the reference signal, which were used to tune the reference signal to bound the estimation error while limiting the actuator demand. 

The derived bounds demonstrated the limitations imposed by the strong directionality of the system, requiring large reference amplitudes for identifying the sensitivity for higher-order modes, even when the dynamics of the reference signal are tuned to the modal closed-loop bandwidth. At the same time, higher-order modes require large input gains to follow the reference signal, conflicting with input magnitude constraints. While these limitations were evaluated using the parameters of the Diamond system, future research could focus on obtaining more general limits that are based on the condition number of the response matrix.

Although the simulations demonstrated that the estimation error remained within the expected bounds, the control inputs were well below the admissible maximum value. Firstly, this resulted from approximating the time-domain constraints in the frequency domain. Secondly, the frequency-domain constraints were enforced using a conservative upper bound, resulting in small input magnitudes for all modes. However, increasing input and reference magnitudes would benefit the signal-to-noise ratio on all modes. Future research could therefore focus developing less conservative bounds for the reference signal, e.g.\ using first principles or predictive control.

Throughout the paper, it was assumed that the plant model is accurate, allowing the MIMO system to be decoupled into sets of SISO systems. Although the Diamond synchrotron is regularly tuned to match its theoretical model, future research could investigate the effect of plant uncertainty, which impacts both the choice of the reference signal and the estimated sensitivity. These results could be further incorporated into a fault detection algorithm, which, in addition to evaluating the algorithm on the real-world system, is subject of future research.

\balance

\FloatBarrier

\section{ACKNOWLEDGMENTS}

This research was supported by the Engineering Undergraduate Research Opportunities Programme (EUROP) scheme of the Department of Engineering Science, University of Oxford, Oxford, UK. Additionally, the authors gratefully acknowledge the contribution of L. Bobb of Diamond Light Source, Oxfordshire, UK.

\balance
\small
\bibliographystyle{IEEEtran}
\bibliography{master_bib_abbrev}

\end{document}